\documentclass[journal]{IEEEtran}

\usepackage{verbatim}
\usepackage{graphicx}
\usepackage[cmex10]{amsmath}
\usepackage{amssymb,amsfonts}
\usepackage{multirow}
\usepackage{theorem}

\newcommand\MYhyperrefoptions
{
	bookmarks=false, 
	bookmarksnumbered=true,
	pdfpagemode={UseOutlines},
	plainpages=false,
	pdfpagelabels=true,
	colorlinks=true,
	linkcolor={black},
	citecolor={black},
	urlcolor={black},
	pdftitle={}, 
	pdfcreator={LaTex},
}

\usepackage{cite}
\usepackage[\MYhyperrefoptions,breaklinks=true,]{hyperref}
\usepackage{breakurl}
\usepackage[tight,footnotesize]{subfigure}
\usepackage{algorithm}
\usepackage{algorithmic}

\usepackage[justification=centering]{caption}  
\usepackage{bbm}

\usepackage{color}
\usepackage{xcolor}
\definecolor{gray}{gray}{0.8} 
\allowdisplaybreaks[4]

\begin{document}

\title{
Balancing Performance and Cost for Two-Hop Cooperative Communications: Stackelberg Game and Distributed Multi-Agent Reinforcement Learning 
}

\author{Yuanzhe~Geng,
	Erwu~Liu,
	Wei~Ni, 
	Rui~Wang, 
	Yan~Liu,
	Hao~Xu, 
    Chen~Cai,
	Abbas~Jamalipour
	\thanks{
		Y. Geng, E. Liu, Y. Liu, and H. Xu are with the College of Electronics and Information Engineering, Tongji University, Shanghai 201804, China, E-mails: yuanzhegeng@tongji.edu.cn, erwu.liu@ieee.org, yanliu2022@tongji.edu.cn, hao.xu@ieee.org.
		
		W. Ni is with Data61, Commonwealth Science and Industrial Research Organization (CSIRO), Marsfield, NSW 2122, Australia, E-mail: wei.ni@data61.csiro.au.
		
		R. Wang is with the College of Electronics and Information Engineering, Tongji University, and is also with the Shanghai Institute of Intelligent Science and Technology, Tongji University, Shanghai 201804, China, E-mail: ruiwang@tongji.edu.cn.
		 
        C. Cai is with the Institute of Carbon Neutrality, College of Environmental Science and Engineering, Tongji University, Shanghai 200092, China, E-mail: caic@tongji.edu.cn.
            
        A. Jamalipour is with the School of Electrical and Information Engineering Faculty of Engineering, The University of Sydney, Sydney, NSW 2006, Australia, Email: a.jamalipour@ieee.org.
		
		Corresponding author: Erwu Liu.
	}
}

\markboth{Submitted to IEEE Transactions on Cognitive Communications and Networking}%
{Shell \MakeLowercase{\textit{et al.}}: }

\maketitle

\begin{abstract}
This paper aims to balance performance and cost in a two-hop wireless cooperative communication network where the source and relays have contradictory optimization goals and make decisions in a distributed manner.
This differs from most existing works that have typically assumed that source and relay nodes follow a schedule created implicitly by a central controller.
We propose that the relays form an alliance in an attempt to maximize the benefit of relaying while the source aims to increase the channel capacity cost-effectively.
To this end, we establish the trade problem as a Stackelberg game, and prove the existence of its equilibrium. 
Another important aspect is that we use multi-agent reinforcement learning (MARL) to approach the equilibrium in a situation where the instantaneous channel state information (CSI) is unavailable, and the source and relays do not have knowledge of each other's goal. 
A multi-agent deep deterministic policy gradient-based framework is designed, where the relay alliance and the source act as agents. 
Experiments demonstrate that the proposed method can obtain an acceptable performance that is close to the game-theoretic equilibrium for all players under time-invariant environments, which considerably outperforms its potential alternatives and is only about 2.9\% away from the optimal solution.

\end{abstract}

\begin{IEEEkeywords}
	Cooperative communication, power control, multi-agent reinforcement learning, Stackelberg game
\end{IEEEkeywords}

\section{Introduction}\label{sect_intro}
Relay-enabled cooperative communication is an effective paradigm for solving problems between resource-limited devices, and has gained much attention for a few decades. 
In order to improve system throughput, a central controller is implicitly assumed to perform optimization operations for the whole communication system.
Accordingly, the optimization problem is defined under the assumption that all sources and relays in such a system will accept and follow this controller's instructions, including relay selection and power allocation. 
Existing methods usually assume the distributions of channel uncertainty and establish the probabilistic model based on instantaneous Channel State Information (CSI). 
Solutions for communication systems can be obtained using optimization tools such as convex optimization \cite{8954800, 9638624, 9194719, 8880523}. 

In practice, it can be difficult for a controller to collect accurate instantaneous CSI in time-varying communication environments. 
Every time the channel conditions change, the resource optimization problem would need to be solved again, incurring overhead and delays.
Several researchers have turned to Reinforcement Learning~(RL) techniques to address the above-mentioned issue and achieved great success~\cite{9655323, 9311792, 10044184, 9046279,DBLP:journals/tccn/KaurTTKPT23}.
RL is one of the most popular machine learning tools in recent years. 
Unlike traditional optimization methods, RL does not rely on prior knowledge or assumptions about the environment during its learning process~\cite{9655323}.
Instead,  RL uses an agent that interacts with the environment. 
The agent learns from experiences and improves its action policy to perform better~\cite{sutton2018reinforcement,9046279}. 
In terms of RL applications in communications, there is no need to rely on assumptions of underlying channels and exact instantaneous CSI. 
RL can directly use observations of communication environments for policy learning. 
RL is a promising tool for many wireless applications. 

This paper delves into cooperative communications, where the source and relays have conflicting utility objectives; i.e., the source wishes to increase its channel capacity with little additional cost, while the relays want to maximize the overall return for assisting the source. 
We establish a Stackelberg game model in support of distributed decision-making, where all relays form an alliance to act as the leader and the source as the follower. 
The participants in the game collaborate by sharing the global CSI of the network.
Yet, they engage in competition to enhance their individual utilities. 
This game-theoretic approach strikes a balance between increasing channel capacity and reducing costs. 

The contributions of this paper are summarized as follows.
\begin{itemize}
    \item 
    We consider a new scenario, where a set of geo-distributed relays form an alliance to help a source node forward signals to an intended destination. The instantaneous CSI of the network may be unavailable. 
    We design a new Stackelberg game between two game players of the source and the relay alliance. 
	
    \item 
    By deriving the optimal solution sets for the game players, we prove the existence of the Nash equilibrium in the Stackelberg game when the instantaneous CSI is available (i.e., the environment is deterministic) and the game players are aware of each other's objective. 

    \item 
    We extend the Stackelberg game to time-varying environments using multi-agent reinforcement learning (MARL), where the instantaneous CSI and competitors' objectives are unavailable to the players. 
    We prove analytically that the utilities of the game-theoretic approach under the availability of the instantaneous CSI and competitors' objectives serve as the upper bounds for those of the MARL-based approach operating with outdated CSI.
    
\end{itemize}
Through extensive simulations, the proposed MARL method demonstrates its superiority to alternative multi-agent methods when the instantaneous CSI is unavailable. Among all learning-based methods that do not require instantaneous CSI, the proposed MARL method outperforms the state-of-the-art single-agent and multi-agent approaches, and achieves the highest utilities and channel capacity.
The MARL method consistently achieves solutions near the Nash equilibrium of the corresponding Stackelberg game operating under instantaneous CSI, with a marginal deviation of approximately 2.9\% from the game-theoretic optimum.

The rest of this paper is organized as follows.
Section \ref{sect works} presents the related works that apply RL to communication systems.
Section \ref{sect model} describes the communication and game model, followed by formulating the Stackelberg game.
Section \ref{game analysis} analyzes the established game and prove the existence of the Nash equilibrium in ideal situations.
Section \ref{sect solution} describes our multi-agent learning framework for solving the Stackelberg game without ideal assumptions and presents our training algorithm in detail.
Section \ref{sect Result} presents the simulation results.
Finally, Section \ref{sect Conclusion} concludes this paper.

\section{Related Work}\label{sect works}
RL has been demonstrated to be an effective tool for solving optimization problems in communications. Many studies have focused on centralized settings, where a central controller is implicitly assumed to act as the RL agent, collecting global information and performing actions for the whole system. For example, 
Su \textit{et. al.} \cite{9137340} employed deep RL, which combined deep neural networks (DNNs) with RL for better generalization, to develop dynamic relay selection schemes for cooperative communication networks.
Tolebi \textit{et al.} \cite{10167871} studied the outage minimization problem in wireless cooperative networks and established a model-free off-policy relay selection model, which was deployed using the deep Q network (DQN).
Nomikos \textit{et al.} \cite{9722918} established a full-duplex relaying model and employed the multi-armed bandits method for learning power control schemes.
The joint relay selection and power control problem was studied in \cite{9568962} and~\cite{9515584}, and the authors designed hierarchical RL architectures to minimize outage probability and maximize total signal-to-noise ratio (SNR), respectively.
These works can cover most of the cooperative communication scenarios. 
However, a central controller may not exist, and multiple controllers are needed for cooperative or non-cooperative learning. The above methods are unsuitable for scenarios where the action policies are learned in a distributed manner.

When there are multiple implicit controllers, a multi-agent game is formed.
As a branch of RL, MARL has recently attracted much attention.
Many researchers have investigated cooperative game problems in cooperative communications, where each device has the same optimization goal. Cooperative games are usually caused by difficulties in obtaining global information. When there is no information exchange between agents, each individual has only partial observations and thus has to perform actions independently~\cite{DBLP:journals/tccn/RahmanKA23}. To achieve better overall performance, researchers usually equip each agent with an RL model, and then, each agent performs cooperative learning to complete a task. 
Lv \textit{et. al.} \cite{9645220} investigated energy-efficient secure transmission and tried to maximize secrecy energy efficiency. 
The authors implemented joint relay and power level selection through multi-agent Q-learning. 
Ortiz \textit{et. al.} \cite{9205292} considered an energy harvesting transmitter and aimed to find distributed transmission policies that maximized the throughput.
Under the assumption that only outdated CSI can be obtained, the authors proposed a Kalman filter-based channel predictor to assist MARL in designing policies.
Gao \textit{et. al.} \cite{9580591} studied the joint trajectory planning of unmanned aerial vehicles (UAV) in a cloud computing architecture, where each UAV acts as an agent and works together to optimize ground users' offloading delay, energy efficiency, as well as obstacle avoidance. 
Then, the authors proposed a game combined with multi-agent DDPG to solve this multi-agent cooperative game.
Hakami \textit{et. al.} \cite{7570172} developed a distributed power control mechanism based on MARL, which optimized the long-run average delay in energy harvesting cooperative relay networks.
In \cite{9508416} and \cite{9322277}, the actor-critic approach was developed in multi-agent situations to maximize the average secrecy rate and perform load balancing.

Cooperative games and cooperative learning are often adopted in cooperative communication.
However, in practice, different participants may have different or even conflicting goals. 
They do not cooperate. 
In such a non-cooperative situation, the optimization and trade-off of system performance and cost in cooperative communications have not been well investigated.
This creates a hybrid model that accommodates cooperation and competition, as well as centralized and decentralized approaches. Unfortunately, the analytical and learning frameworks from prior studies may not readily apply in such scenarios due to alterations in the agent relationships and interaction processes.

\section{System Model and Problem Formulation}\label{sect model}
In this section, we first introduce the model of our considered two-hop cooperative communication network. 
Then, we establish a Stackelberg game for this communication model.

\subsection{Communication Model}\label{subsect comm model}
\begin{figure} 
	\centering
	\includegraphics[scale=0.555]{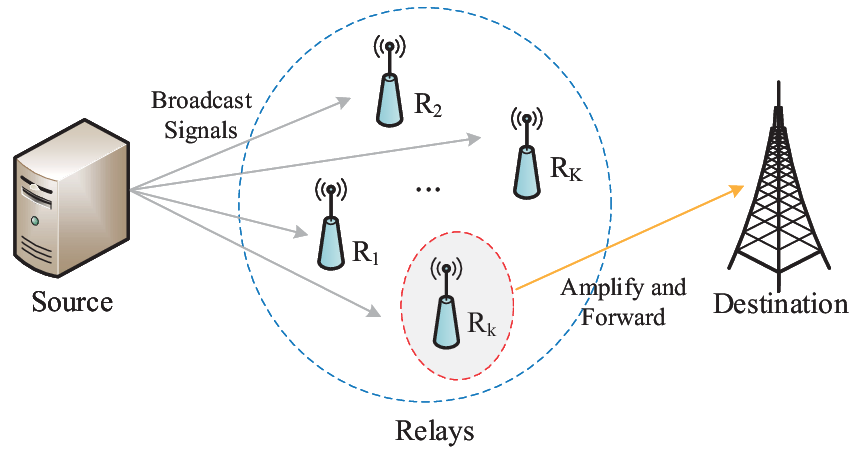}
	\caption{\small The illustration of a two-hop relay-enabled cooperative network.}
	\label{relay model}
\end{figure}

As shown in Fig. \ref{relay model}, there is a source $S$, a destination $D$, and a group of $K$ relays in the considered two-hop wireless relay network.
The source communicates with the destination with the assistance of a selected relay, where the amplify-and-forward (AF) protocol is adopted.

Consider a typical half-duplex mode.
The communication from the source $S$ to the destination $D$ via relay $R_k, 1 \leq k \leq K$, takes two time slots.
In the first time slot, the source transmits its signal.
Assume that the direct link is obstructed between the source and destination.
Only relays receive this transmission.
The received signal at relay $R_k$ is written as 
\begin{equation} y_{sk}(t)=\sqrt{P_s} h_{sk}(t)x(t)+n_k(t), \end{equation}
where $P_s$ represents the transmit power of the source; $x(t)$ represents data symbol; 
$h_{sk}(t)$ represents the channel gain between the source and relay $R_k$, which is assumed to be a complex Gaussian random variable with zero mean and variance $\sigma_{sk}^2$; and $n_k(t)$ is the complex Gaussian noise at relay $R_k$, and it has zero mean and variance $\sigma_k^2$.
Further, to characterize the temporal correlation between consecutive time slots in each channel, we employ the following Gaussian Markov block fading autoregressive model \cite{5710995, 9568962}: 
	\begin{equation} \label{channel state} h_{ij}(t)=\rho h_{ij}(t-1)+\sqrt{1-\rho^2} \zeta(t),  \end{equation}
	where the subscript ``$_{i, j}$'' indicates two different nodes, $\rho$ is the normalized channel correlation coefficient, and $\zeta(t)\sim\mathcal{CN}(0,\sigma_j^2)$ is the random change uncorrelated with $h_{ij}(t)$.

In the second time slot, a selected relay $R_k$ amplifies and forwards its detected signal to the destination.
Then, the received signal at the destination can be written as
\begin{equation}
y_{kd}(t)=\sqrt{P_k} h_{kd}(t)x_{sk}(t)+ n_d(t),
\end{equation}
where $P_k$ is the transmit power of the relay $R_k$, and $x_{sk}(t)=y_{sk}(t)/\|y_{sk}(t)\|$. 
Similarly, $h_{kd}(t)$ represents the channel gain between relay $R_k$ and the destination, which is also assumed to be a complex Gaussian random variable with zero mean and variance $\sigma_{kd}^2$. 
$n_d(t)$ is the complex Gaussian noise at the destination with zero mean and variance $\sigma_d^2$.
Then, we have 
\begin{equation}\begin{aligned}
y_{kd}(t)=
&\frac{\sqrt{P_k}\sqrt{P_s}h_{kd}(t)h_{sk}(t)x(t)}{\sqrt{P_s{\|h_{sk}(t)\|}^2+\sigma_k^2}} 
\\ & + \frac{\sqrt{P_k} h_{kd}(t)n_k(t)}{\sqrt{P_s{\|h_{sk}(t)\|}^2+\sigma_k^2}} 
+ n_d(t),
\end{aligned}\end{equation}
The output SNR at the destination can be written as \cite{9137340, 5710995}
\begin{equation}
\gamma = \frac{P_s P_k {\|h_{sk}(t)\|}^2 {\|h_{kd}(t)\|}^2}{P_k\|h_{kd}(t)\|^2 \sigma_k^2 + \sigma_d^2(P_s\|h_{sk}(t)\|^2+\sigma_k^2)}.
\end{equation}
Accordingly, we have the instantaneous end-to-end channel capacity as follows.
\begin{equation}\begin{aligned}
I
=\frac{1}{2}\log_2(1+\gamma)
=\frac{1}{2}\log_2(1+\frac{P_k \gamma_{sk}}{P_k + G_k}),
\end{aligned}\end{equation}
where 
$\gamma_{sk}=\frac{P_s\|h_{sk}(t)\|^2}{\sigma_k^2}$ 
and $G_k=\frac{\sigma_d^2(P_s\|h_{sk}(t)\|^2+\sigma_k^2)}{\sigma_k^2\|h_{kd}(t)\|^2}$ 
are defined for conciseness.

\subsection{Stackelberg Game Model}\label{subsect game model}
The relays need to consume their power to assist cooperative communications. 
The source needs to pay for its selected amount of power. 
We set up a Stackelberg game that allows participants/agents to carry out distributed decision-making based on their observations, e.g., CSI.

On one side of the Stackelberg game is the relay alliance formed by all relays, as the relays may have the same ownership, which helps prevent ill competition. 
The alliance operates centrally.
The alliance nominates a relay and charges a price $C_k$ per power unit. 
As a result, we can define the following utility function:
\begin{equation}\label{utility relay}
U^r(C_k, P_k) = C_k P_k, 
\end{equation}
The utility accounts for the total revenue of the relay alliance, which is the goal of the alliance to maximize. 
Thus, the optimization problem of the relays can be formulated as
\begin{equation} \begin{aligned} \label{problem relay}
&\max\limits_{k, C_k}\ U^r(C_k, P_k), \\
\textit{s.t.}
\ C_{min} &\leq C_k \leq C_{max}, 1 \leq k \leq K,
\end{aligned}\end{equation}
where the first constraint gives the bounds of the price, and the second constraint specifies the index range of relays.

The source is on the other side of the Stackelberg game.
Based on the given power price $C_k$, the source determines the power amount $P_k$ that needs to be purchased from the relay alliance.
The source aims to maximize its utility function as
\begin{equation}\begin{aligned}\label{utility source}
U^s(P_k, C_k)
=\frac{1}{2}\log_2(1+\frac{P_k \gamma_{sk}}{P_k + G_k}) - \alpha C_k P_k,
\end{aligned}\end{equation}
where $\alpha$ is a weighting coefficient. 
Similarly, the optimization problem of the source can be written as
\begin{equation}\begin{aligned} \label{problem source}
&\max\limits_{P_k}\ U^s(P_k, C_k), \\
\textit{s.t.}
\ &P_{min} \leq P_k \leq P_{max},
\end{aligned}\end{equation}
where the constraint gives the bound of the power amount. 
The destination is captured implicitly in the utility of the source in (\ref{utility source}). 
When trading with a selected relay $k$, the objective of the source is to increase its end-to-end channel capacity with less economic cost. 
Since the relays assist the communication between the source and the destination, the destination has the same goal as the source and is captured.

The relay alliance and the source jointly form a Stackberg game. 
Both serve as game players, where the relay alliance is the leader and the source is the follower.
This is because the relay alliance holds an inherent advantage over the source, stemming from factors, such as its position in the network, its ability to amplify signals, and its knowledge about channel conditions (from the source to the relays). By assuming the role of the leader, the relay alliance can exploit this advantage to influence the source's behavior and optimize the overall performance of the considered communication system. In particular, the relay alliance can shape the communication process in a way that maximizes the likelihood of successful data transmission.

In the ensuing section, we analyze the best responses of the follower and the leader by considering each other’s strategies, and prove the existence of the Nash equilibrium in each stage of this Stackelberg game.

\section{Game Analysis and Solution}\label{game analysis}
In this section, we propose the optimal solution for the Stackelberg game under ideal situations where the game players' goals are known and the instantaneous CSI is available. 
To do this, we first analyze the concavity of the game players' utility functions, and then derive their best responses. 
At last, we obtain the optimal solution for each player at the Nash equilibrium of the Stackelberg game. 
Apart from the relay alliance (as described in Section~\ref{subsect game model}), we also analyze a setting of competitive relays. 
We show that the relay alliance under the cooperative setting can bring larger revenue than each individual relay under the competitive setting.

\subsection{Analysis of Relay Alliance}\label{subsect analysis cooperative}
We start by analyzing the best response of the source (i.e., the follower). 
Given the selected relay $R_k$ and the price $C_k$ announced by the relay alliance  (i.e., the leader), the source needs to determine the transmit power to be purchased from the relay alliance. 
Usually, the utility $U^s$ is higher with a higher transmit power $P_k$. 
Nevertheless, when $P_k$ is too large, the utility decreases due to high cost. 
Therefore, the source must choose the proper transmit power of the relay.

\vspace{3 mm}
\textit{\textbf{Proposition 1. }
Given the price per unit of power $C_k$, the utility of the source is strictly concave with respect to the transmit power of the relay, $P_k$.}
\vspace{3 mm}

\textit{Proof. }
The second derivative of utility $U^s$ with respect to $P_k$ is given by 
\begin{equation} \begin{aligned}
\frac{\partial^2 U^s}{\partial P_k^2}
=-\frac{\gamma_{sk} G_k\big((2\gamma_{sk}+2)P_k+(\gamma_{sk}+2)G_k\big)}
{2\log 2(P_k+G_k)^2\big((\gamma_{sk}+1)P_k+G_k\big)^2},
\end{aligned}\end{equation}
which is negative since $\gamma_{sk}>0$ and $G_k>0$.
The concavity of $U^s$ is proved. 
$\hfill \square$
\vspace{3 mm}

Based on \textbf{Proposition 1}, the best response of the source can be obtained by setting the first derivative of the utility $U^s$ to zero, 
and then solving the equation for $P_k$, i.e., 
\begin{equation} \begin{aligned}
\frac{\partial U^s}{\partial P_k}
=\frac{\gamma_{sk}G_k}{2\log 2(P_k+G_k)(\gamma_{sk}P_k+P_k+G_k)}-\alpha C_k=0,
\end{aligned}\end{equation}
with the solution of 
\begin{equation}\begin{aligned} \label{solution_pk}
P_k^*&=
\big(
\sqrt{\log 2}\sqrt{\log 2 \alpha^2 \gamma_{sk}^2 G_k^2 C_k^2 + (2\gamma_{sk}^2+2\gamma_{sk})\alpha G_k C_k} \\
&-\log 2 (\gamma_{sk}+2) \alpha G_k C_k
\big)
/
\big(
2\log 2(\gamma_{sk}+1) \alpha C_k
\big).
\end{aligned}\end{equation}

The best response $P_k^*$ can be further written as 
\begin{equation} \begin{aligned} 
P_k^* 
= &\frac{\sqrt{\log 2}\sqrt{\log 2 \alpha^2 \gamma_{sk}^2 G_k^2 C_k^2 + (2\gamma_{sk}^2+2\gamma_{sk})\alpha G_k C_k}
}
{2\log 2 (\gamma_{sk} + 1) \alpha C_k} \\ 
&- \frac{\log 2 (\gamma_{sk} + 2) \alpha G_k C_k}{2\log 2 (\gamma_{sk} + 1) \alpha C_k} \\
= &\sqrt{CO_1 + CO_2 / C_k} - CO_3,
\end{aligned}\end{equation}
where 
\begin{equation} \begin{aligned} 
&CO_1 = \frac{(\log 2)^2 \gamma_{sk}^2 G_k^2}{(2\log 2\gamma_{sk}+2\log 2)^2}; \\ 
&CO_2 = \frac{(2\gamma_{sk}^2 + 2\gamma_{sk})G_k}{(2\log 2\gamma_{sk}+2\log 2)^2 \alpha}; \\
&CO_3 = \frac{(\log 2 \gamma_{sk}+2\log 2) G_k}{2\log 2\gamma_{sk}+2\log 2}.
\end{aligned}\end{equation}

Clearly, $P_k^*$ decreases as $C_k$ increases ($C_k>0$). 
On the other hand, $P_k^*$ is also bounded; i.e., $P_{min} \leq P_k^* \leq P_{max}$.
Let $\sqrt{CO_1 + CO_2 / C^0} - CO_3 = P_{max}$, and $\sqrt{CO_1 + CO_2 / C^1} - CO_3 = P_{min}$.
Then, we have 
\begin{equation}\begin{aligned} 
&C^0 = \frac{CO_2}{(P_{max}+CO_3)^2 - CO_1}; \\
&C^1 = \frac{CO_2}{(P_{min}+CO_3)^2 - CO_1}.
\end{aligned}\end{equation}
With the given $C_k$, the source's optimal decision is one of $P_k^*$, $P_{max}$, and $P_{min}$, as given by 
\begin{equation}\label{source decision}
P_k = \left\{
\begin{aligned}
P_{max}, & \ \text{if} \ C_k < C^0 ;\\
P_{min}, & \ \text{if} \ C_k > C^1; \\
P_k^*, & \ \text{otherwise}.
\end{aligned}
\right.
\end{equation}

Next, we analyze the best response of the relay alliance (i.e., the leader), or more specifically, relay $k$ that is selected by the relay alliance to serve the source. 
The relay alliance sets its unit power price $C_k$ to maximize its revenue.
It is clear that if the unit price is too high, the source will reduce the power that it purchases, and the revenue received by the alliance can decrease.
As a result, although the leader can take action first to maximize its utility, it still needs to take the follower's reaction into consideration.

Suppose the players know each other's goal.
Then, for each possible candidate $R_k$ in the alliance, we can substitute the follower's best response into the leader's utility. 
It follows that 
\begin{equation}\begin{aligned}
U^r(C_k, P_k) = C_k P_k^* .
\end{aligned}\end{equation}

\vspace{3 mm}
\textit{\textbf{Proposition 2. }
Given the transmit power $P_k$ derived by the follower's best response, the utility of the leader is strictly concave with respect to the unit power price $C_k$.}
\vspace{3 mm}

\textit{Proof. }
The second derivative of utility $U^r$ with respect to $C_k$ is given by 
\begin{equation} \begin{aligned}
\frac{\partial^2 U^r}{\partial C_k^2}
=\frac{-\gamma_{sk}^2(\gamma_{sk}+1) \alpha G_k^2}
{2\sqrt{\log 2}\big(\log 2\alpha^2\gamma_{sk}^2 G_k^2 C_k^2+(2\gamma_{sk}^2+2\gamma_{sk})\alpha G_kC_k \big)^{\frac{3}{2}}}.
\end{aligned}\end{equation}
Clearly, $\frac{\partial^2 U^r}{\partial C_k^2}<0$.
The concavity of $U^r$ with respect to $C_k$ is proved, given the follower's best response $P_k$.
$\hfill \square$
\vspace{3 mm}

According to \textbf{Proposition 2}, the optimal power price $C_k$ can be uniquely obtained by setting the first derivative of utility $U^r$ to zero, and then solving the equation, i.e., 
\begin{equation}
\frac{\partial U^r}{\partial C_k}
=P_k^*+C_k\frac{d P_k^*}{d C_k}=0.
\end{equation}
After mathematical manipulations, the best response of the relay alliance can be given by 
\begin{equation}
C_k^*=
\big|
\frac{B_k D_k \sqrt{D_k^2 - \log 2 A_k} - B_k D_k^2 + \log 2 A_k B_k}
{2 A_k D_k^2 - 2\log 2 A_k^2}
\big|,
\end{equation}
where 
$A_k=\log 2 \alpha^2 \gamma_{sk}^2 G_k^2$,  
$B_k=2 \alpha (\gamma_{sk}^2+\gamma_{sk}) G_k$,  
and $D_k=\alpha (\log 2 \gamma_{sk} + 2\log 2) G_k$.  

It is possible that the optimal decision of the relay alliance, $C_k^*$, results in the transmit power that the source wants from the relay exceeding its threshold. 
In this sense, we need to consider that the leader's price may lead to the derived follower's optimal power exceeding the threshold. 
Specifically, the utility function of the relay alliance is $U^r(C_k, P_k) = C_k P_k$, which is continuous in $C_k > 0$. 
When the source's optimal decision is set to $P_{max}$ or $P_{min}$, we can find that $U^r$ is a monotonically increasing function in $0 < C_k \leq C^0$ and $C_k \geq C^1$.
The monotonicity of function $U^r$ in $C^0 \leq C_k \leq C^1$ needs to be discussed case-by-case by comparing the best response $C_k^*$ with $C^0$ and $C^1$: 
\begin{enumerate}
    \item If $C_k^* < C^0$, then $U^r$ decreases monotonically in $C^0 \leq C_k \leq C^1$.
    \item If $C_k^* > C^1$, then $U^r$ increases monotonically in $C^0 \leq C_k \leq C^1$.
    \item Otherwise, $U^r$ increases monotonically in $C^0 \leq C_k \leq C_k^*$, and decreases monotonically in $C_k^* \leq C_k \leq C^1$.
\end{enumerate}
As a result, the maximum $U^r$ can be taken at $C_k = C^0$ 
but can never be taken at $C_k = C^1$ unless $C^1=C_{max}$.
With the constraint of $ C_{min} \leq C_k \leq C_{max}$, the relay alliance's optimal decision is taken from the following set: 
    \begin{equation}\label{price optimal set cooperative}
        \textbf{C}_k^{opt} = \{ C_{min}, C_{max}, C_k^*, C^0 \}.
    \end{equation}

Nominating different relays can result in different maximum utilities of the relay alliance. 
By comparing the different possible utilities and finding the maximum, the relay alliance can determine the best relay $R_k$ and its corresponding price $C_k$ from $\textbf{C}_k^{opt}$. 
Accordingly, the optimal decision of the source can be determined through (\ref{source decision}). 
The existence of the Nash equilibrium can be confirmed~\cite{bacsar1998dynamic, fiez2020implicit}.
As a result, both the relay alliance and the source can achieve their maximum utilities in the Stackelberg game.

\subsection{Analysis of Competitive Relays}\label{subsect analysis competitive}
Competitive relays have been widely considered in two-hop communication networks~\cite{8949705, 7467570, 10283689}. 
When relays are competitive, each relay $k$ acts as an independent agent to maximize its own revenue in (\ref{utility relay}) by determining its price $C_k$. 
Unlike the relay alliance (as considered in Section \ref{subsect analysis cooperative}), the source needs to select a relay. 
Based on the given price of each relay, the source selects a relay $k$ and specifies its transmit power $P_k$, to maximize its utility function in (\ref{utility source}). 

All relays need to compete to get selected by the source.
Obviously, function $U^s$ is continuous and decreases monotonically in $C_k > 0$.
To provide the source with a higher utility and hence get selected more likely, relay $k$ reduces its price $C_k$ to increase $U^s(P_k^*, C_k)$.
Only one relay wins this competition: 
\begin{equation}
	k^* = \arg \max_i U^s(P_i^*, C_{i,min}), i \in \{1, 2, \cdots, K\},
\end{equation}
where $k^*$ is the index of the winner relay, and $P_i^*$ denotes the source's optimal decision on the transmit power of the $i$-th relay that maximizes $U^s$ given the price $C_{i,min}$ of the relay.

The lower bound of the source's optimal utility can be written as
\begin{equation}\label{lower bound of us}
	\tilde{U}^s = \max_i U^s(P_i^*, C_{i,min}), i \in \{1,\cdots,k-1,k+1,\cdots,K \}.
\end{equation}
By solving the following equation with respect to $C_k$,
\begin{equation}
	U^s(P_k^*, C_k) - \tilde{U}^s = 0,
\end{equation}
we can obtain the unique solution, denoted as $C_{k,max}^{game}$, which is an upper bound for the price of relay $k$, i.e., $C_k < C_{k,max}^{game}$.

\vspace{3 mm}
\textit{\textbf{Proposition 3. }
If the winner relay $k$ announces its price $C_k$, $C_{k,min} \leq C_k \leq C_{k,max}$ and $C_k < C_{k,max}^{game}$, then the source's optimal utility $U^s(P_k^*, C_k)$ is higher than the lower bound $\tilde{U}^s$ defined in (\ref{lower bound of us}), and the source always chooses relay $k$.}
\vspace{3 mm}

After the winner relay $k$ is determined, the relay and the source form a Stackelberg game, which can be analyzed in the same way as the relay alliance. 
To this end, the source's optimal decision is still among $P_k^*$, $P_{max}$, and $P_{min}$, as in (\ref{source decision}). 
With the additional upper bound for relay $k$'s price, the optimal decision of the relay is taken from the following set: 
    \begin{equation}\label{price optimal set competitive}
        \tilde{\textbf{C}}_k^{opt} = \{ C_{k,min}, C_{k,max}, C_k^*, C^0, C_{k,max}^{game}-\epsilon \},
    \end{equation}
where $\epsilon$ is an arbitrarily small, positive, real value. 
Based on the above analysis, the winner relay determines its price from $\tilde{\textbf{C}}_k^{opt}$ by finding the maximum, then the source's optimal decision on the transmit power, $P_k$, is determined through (\ref{source decision}).
The existence of the Nash equilibrium can be confirmed under the competitive setting.
Then, we put forth the following proposition.

\vspace{3 mm}
\textit{\textbf{Proposition 4. }
The relay alliance can bring larger revenue $U^r$ than competitive relays.}
\vspace{3 mm}

\textit{Proof. }
The winner relay $k$ has a larger set of possible optimal decisions, including an additional element, $C_{k,max}^{game}-\epsilon$, under the competitive setting than under the relay alliance. 
Nevertheless, the optimal revenue in the cooperative setting is larger than in the competitive setting. 
This is because, with the additional upper bound under the competitive setting (i.e., $C_k < C_{k,max}^{game}$) on $\tilde{\textbf{C}}_k^{opt}$, we have 
\begin{itemize}
    \item 
    If $C_{k,max}^{game} > C_{k,max}$, the possible optimal decision sets of the winner relay $k$ and the relay alliance are the same: 
    \begin{equation}
        \tilde{\textbf{C}}_k^{opt} \setminus \{C_{k,max}^{game}-\epsilon\} = \textbf{C}_k^{opt}.
    \end{equation}
    In this case, the optimal revenues under the cooperative and competitive settings are equal.

    \item 
    If $ C_{k,min} < C_{k,max}^{game} \leq C_{k,max}$, then 
    $C_{k,max}$ must not be selected from $\tilde{\textbf{C}}_k^{opt}$ because $C_{k,max}^{game}-\epsilon < C_{k,max}$. 
    Further, $C^0$ cannot be selected if $C_{k,max}^{game} \leq C^0$; 
    $C_k^*$ cannot be selected if $C_{k,max}^{game} \leq C_k^*$. 
    Let $\tilde{\textbf{C}}_k^{ex}$ collects these elements precluded from $\tilde{\textbf{C}}_k^{opt}$. 
    The optimal decision under the competitive setting should be taken from  
       $\tilde{\textbf{C}}_k^{opt} \setminus \{C_{k,max}\} \setminus \tilde{\textbf{C}}_k^{ex}$. 
    Since the maximum $U^r$ cannot be taken at $C_k = C_{k,max}^{game}$, we have 
    \begin{align}
    	U^{r}(C_{k,max}^{game}) < \max& \{U^{r}(C^0), U^{r}(C_k^*), \nonumber\\
            &U^{r}(C_{k,max}), U^{r}(C_{k,min}) \}, 
    \end{align}
    and, consequently,  
    \begin{equation}
        \max_{C_k \in 
        \tilde{\textbf{C}}_k^{opt} \setminus \{C_{k,max}\} \setminus \tilde{\textbf{C}}_k^{ex}
        } U^r 
        \leq \max_{C_k \in 
        \textbf{C}_k^{opt}
        } U^r.
    \end{equation}
    The optimal revenue in the cooperative setting is larger than, or equal to, that in the competitive setting. 
\end{itemize}
This proof is complete. $\hfill \square$

\vspace{3 mm}

As a result, the relays preferably cooperate and form an alliance to compete with the source to increase their revenue. 
This confirms the benefit of forming the relay alliance and the Stackelberg game between the source and the relay alliance.

\section{MARL for Stackelberg Game}\label{sect solution}
For the Stackelberg game formed by (\ref{problem relay}) and (\ref{problem source}), the analysis in Section-\ref{game analysis} indicates the existence of the optimal solution in the ideal situation where the players know each other's goal and strategy. 
Specifically, both the source and relay alliance make decisions based on the instantaneous CSI under the assumption that the players' utilities are known to all players. 
In practice, however, the channels change rapidly, and it can be difficult to obtain the CSI immediately. 
As the competitors in the same game, the relay alliance and source may not reveal their intents and actions. Each player only knows its own utility goal. 
In this case, we further propose an MARL-based approach in time-varying environments where the instantaneous CSI and competitors’ objectives are unavailable, and the source and relay alliance can only make decisions based on outdated CSI.

In this section, we model the learning process as a Markov decision process (MDP) and propose a DDPG-based multi-agent learning scheme to solve the game.
The two players (i.e., the source and the relay alliance) are two RL agents. 
The agents obtain outdated CSI.
The relay alliance agent (hereinafter referred to as ``relay agent" for short) acts before the source agent, selects the relay, and sets the price according to outdated CSI. 
Then, the source agent controls the transmit power, according to both outdated CSI and the behavior of the relay agent.
Both agents receive their reward from the environment, and a new round of the game begins.

\subsection{Markov Decision Process}\label{subsect MDP}
An MDP, which can be denoted as $<\mathcal{E}, \mathcal{S}, \mathcal{A}, \mathcal{R}, p>$, consists of an environment $\mathcal{E}$, a state space $\mathcal{S}$, an action space $\mathcal{A}$, a reward space $\mathcal{S}\times\mathcal{A}\to\mathcal{R}$, and a transition probability $p$. 
At each discrete time step $t$, the RL agent first observes the current state $s_t\in\mathcal{S}_t$.
Then, it selects an action $a_t\in\mathcal{A}_t$, according to its policy $\pi$ that maps states to a probability distribution over actions. 
After executing action $a_t$, the agent receives a scalar reward $r_t \in \mathcal{R}_t$ from the environment $\mathcal{E}$, and observes the next state $s_{t+1}$ that follows the transition probability $p(s_{t+1}|s_t,a_t)$. 
The transition probability of the state is unknown to the agent.
As a result, the learning stage is a process of trial and error to obtain the optimal policy that maximizes the agent's accumulated reward $R_t = \sum\nolimits_{i=t}^T\gamma^{i-t}r_i$, where variable $T$ denotes the total number of steps, and $\gamma\in[0,1]$ denotes a discount factor.

\begin{figure} 
	\centering
	\includegraphics[scale=0.58]{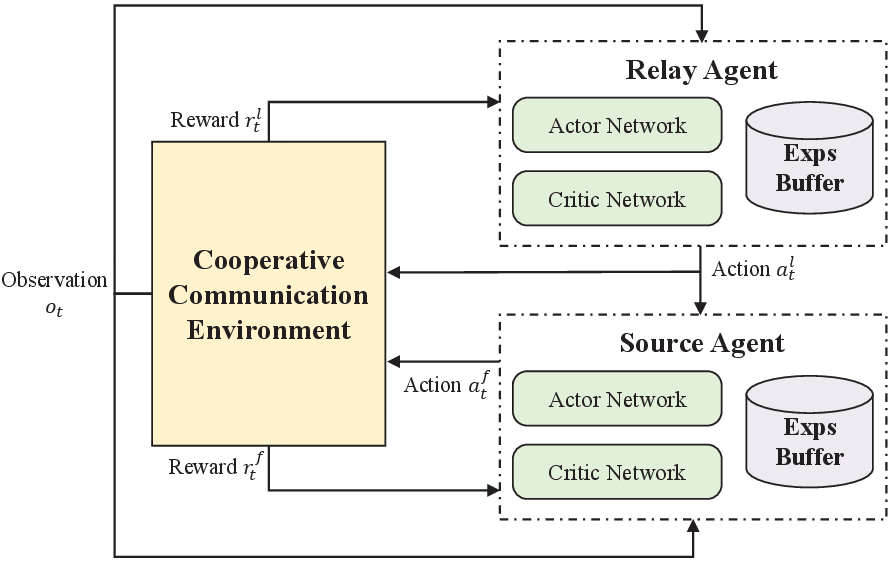}
	\caption{Multi-agent RL interaction framework for the Stackelberg game.}
	\label{RL framework}
\end{figure}

According to the communication and game models described in Section-\ref{sect model}, we design the RL system for the investigated multi-agent environment, as shown in Fig.~\ref{RL framework}. 
Considering rapidly changing channels and subsequently inevitable delays in the acquisition of the CSI,  only outdated CSI is available. 
For this reason, each agent sets the CSI of the previous time slots as its observation at the current time slot.
Accordingly, the full observation space of the agents is the union of different wireless channel states, i.e.,
\begin{equation} \label{obveservation} 
\mathcal{O}_t \triangleq [h_{sk}(t-1), h_{kd}(t-1)], \, k=1,\cdots,K.
\end{equation}

Since there are two independent RL agents in the environment, we define their system variables separately. 
The system variables of the \textbf{relay agent} are set as follows.
\begin{itemize}
\item \textbf {System State: }
The relay agent acts before the source agent. Thus, it directly employs the observation as its system space. 
Therefore, we have 
\begin{equation} \label{state_space_l} 
    \mathcal{S}_t^l = \mathcal{O}_t \triangleq [h_{sk}(t-1), h_{kd}(t-1)], 
\end{equation}
where the superscript $``l"$ in variable $\mathcal{S}_t^l$ means the leader in the Stackelberg game.
	
\item \textbf {Action: }
In each time slot, the relay agent needs to choose an optimal relay to assist the source in forwarding the signal, and set the price per unit of power.  
Therefore, the action of the relay agent is defined as 
\begin{equation}\label{action_space_l} 
    \mathcal{A}_t^l \triangleq [a^{R}(t), a^{C}(t)], 
\end{equation}
where $a^{R}(t) \in \{1,2,\dots,K\}$ is the index of the selected relay, and $a^{C}(t) \in [C_{min}, C_{max}]$ is the power price.

\item \textbf {Reward Function: }
According to the utility function of the relay alliance in (\ref{utility relay}), we define the following reward function for the relay agent: 
\begin{equation}\label{reward_l} r_t^l = \beta U^r, \end{equation}
where $\beta$ is a scaling parameter, which is introduced to make the agents' rewards in the same order of magnitude to enhance the stability of training.
\end{itemize}

The \textbf{source agent} has its own system state, action, and reward function.
The settings are as follows.
\begin{itemize}
\item \textbf {System State: }
Before the source agent performs actions, the actions of the relay agent are already known. 
Therefore, the state space of the source agent includes both the initial observation of the communication environment and the relay agent's behavior, as follows.
\begin{equation}\begin{aligned}\label{state_space_f} 
    \mathcal{S}_t^f = [\mathcal{O}_t, \mathcal{A}_t^l] 
    \triangleq [h_{sk}(t-1), h_{kd}(t-1), a^{R}(t), a^{C}(t)], 
\end{aligned}
\end{equation}
where the superscript $``f"$ indicates the follower in the Stackelberg game.
	
\item \textbf {Action: }
According to the formed Stackelberg game, in each time slot, the source only needs to determine the amount of power to purchase from the selected relay.
Therefore, the action of the source can be defined as 
\begin{equation}\label{action_space_f} \mathcal{A}_t^f \triangleq [a^{P}(t)], \end{equation}
where $a^{P}(t) \in [P_{min}, P_{max}]$.
	 
\item \textbf {Reward Function: }
We can employ the utility function of the source in (\ref{utility source}) as the source's reward function: 
\begin{equation}\label{reward_f} r_t^f = U^s. \end{equation}
\end{itemize}

\subsection{DDPG Based MARL Solution}\label{subsect DDPG}
\begin{figure} 
	\centering
	\includegraphics[scale=0.65]{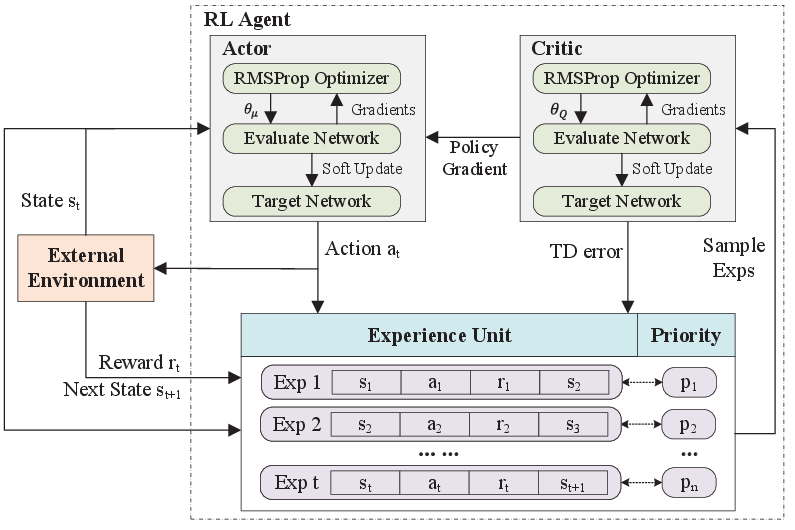}
	\caption{Internal architecture of a single DDPG agent with prioritized experience buffer.}
	\label{DDPG}
\end{figure}

The DDPG method has been proven effective in solving problems with continuous action spaces \cite{DDPG, 9444840, 9453731}.
To achieve optimal policies for both the relay agent and source agent, we adapt the state-of-art DDPG method into the investigated Stackelberg game, and design a multi-agent framework, as depicted in Fig. \ref{RL framework}.
We assume that the source agent and relay agent have the same architecture.
The architecture of each agent is depicted in Fig. \ref{DDPG}.

As depicted in Fig. \ref{DDPG}, each agent consists of two parts, namely actor and critic, and each part has two DNNs.
One DNN works as an evaluate network, and the other one is a copy of the previous version of the evaluate network, which is also known as the target network. 
To avoid instability in different scale states and improve the sampling efficiency during policy learning, we employ an independently prioritized experience buffer for each agent.
The buffer stores the agent's experience $e_t=\{s_t, a_t, r_t, s_{t+1}\}$ after each interaction, and each experience unit has a priority evaluation $p_t = |\delta_t| + \epsilon$ that indicates its importance. 
The small positive constant $\epsilon$ ensures that each unit has a non-zero probability to be
sampled even if its temporal difference (TD) error $\delta_t$ is zero.

For illustration convenience, we take the source agent as an example to describe the architecture design. 

\textbf{Critic:} 
The critic estimates the value of each action by employing a DNN with parameter $\theta_Q^f$. 
Define the action-value function $Q(s_t, a_t; \theta)=\mathbb{E}[R_t|s_t,a_t]$ to represent the expected return after performing action $a_t$ in state $s_t$. 
The critic has the following loss function: 
\begin{equation} \label{critic loss f}
L(\theta_Q^f)=
\mathbb{E}_{e_t^f\sim\mathcal{B}^f}\Big[w_t^f \cdot \big( 
	\delta_t^f(s_{t+1}^f, a_{t+1}^f; \theta_Q^f)
\big)^2],
\end{equation}
where $\delta_t^f(s_{t+1}^f, a_{t+1}^f; \theta_Q^f) = r_t^f + \gamma \max\nolimits_{a_{t+1}^f}Q(s_{t+1}^f, a_{t+1}^f; \bar{\theta}^f_Q) - Q(s_t^f, a_t^f; \theta_Q^f)$ is the TD error, 
$\bar{\theta}^f_Q$ is the old parameters in the critic's target network; 
and $w_t^f$ is a normalized importance-sampling weight, which is based on the priority, i.e., 
\begin{equation}
w_i^f = \frac{(|\mathcal{B}^f| \cdot p_i^f)^\kappa}{\max\nolimits_{j<t} w_j^f},
\end{equation}
where $|\mathcal{B}^f|$ is the size of the experience buffer $\mathcal{B}^f$, and $\kappa$ is an exponent. $0<\kappa<1$.

During the learning process, the old parameters are softly replaced following $\bar{\theta}^f_Q \leftarrow \tau\theta_Q^f + (1-\tau)\bar{\theta}^f_Q$ with $\tau \ll 1$.
Then, the parameters of the evaluate network are updated using RMSProp optimization, as given by 
\begin{equation} \begin{aligned} \label{critic update f}
\theta_Q^f \leftarrow 
\theta_Q^f - \eta_Q^f w_t^f \delta_t^f(s_{t+1}^f, a_{t+1}^f; \theta_Q^f) \nabla_{\theta_Q^f}Q(s_t^f, a_t^f; \theta_Q^f),
\end{aligned}\end{equation}
where $\eta_Q^f$ is the learning step size in the critic.

\begin{algorithm} 
    \caption{Multi-Agent DDPG for Stackelberg Competition of Cooperative Communication.}
	\label{algo MADDPG}
	\begin{algorithmic}[1]
		\STATE Initialize the experience replay buffers $\mathcal{B}^f, \mathcal{B}^l$.
		\STATE Initialize the follower's evaluate network $\theta_Q^f, \theta_\mu^f$, and the leader's evaluate network $\theta_Q^l, \theta_\mu^l$.
		\STATE Initialize the target networks with $\bar{\theta}_Q^f=\theta_Q^f, \bar{\theta}_\mu^f=\theta_\mu^f$, and $\bar{\theta}_Q^l=\theta_Q^l, \bar{\theta}_\mu^l=\theta_\mu^l$.
		\FOR{episode $t^{epi}=1,\dots,u_{max}^{epi}$}
		\STATE Initialize the communication environment, and obtain initial observation of the environment $o_0$.
		\STATE Initialize the random process $\Delta\mu^f, \Delta\mu^l$ as noises for the follower and the leader.
		\FOR{time slot $t=1,2,\dots,t_{max}$}
		\STATE The leader chooses an action $a_t^l=\mu(s_t^l;\theta_\mu^l)+\Delta\mu_t^l$ to determine the selected relay and price per power unit.
		\STATE The follower observes the leader's action $a_t^l$ and constructs its state $s_t^f=(o_t, a_t^l)$, then it chooses an action $a_t^f=\mu(s_t^f;\theta_\mu^f)+\Delta\mu_t^f$ to determine the amount of power that buys from the leader.
		\STATE Both the leader and the follower execute actions $a_t^l, a_t^f$, then receive their own rewards $r_t^f, r_t^l$ and obtain the next observation of the environment $o_{t+1}$.
		\FOR{both the leader and the follower}
		\STATE Collect and save experience tuple $e_t$ in its buffer $\mathcal{B}$, then sample a mini-batch of transitions $(s_j, a_j, r_j, s_{j+1})$. 
		\STATE Minimize its critic's loss in (\ref{critic loss f}), and update the evaluate network of critic according to (\ref{critic update f}).
		\STATE Calculate the sampled policy gradient in (\ref{actor gradient f}), and update the evaluate network of the actor, according to (\ref{actor update f}). 
		\STATE Update parameters of its target networks by $\bar{\theta}_Q \leftarrow \tau\theta_Q + (1-\tau)\bar{\theta}_Q$ and $\bar{\theta}_\mu \leftarrow \tau\theta_\mu + (1-\tau)\bar{\theta}_\mu$.
		\ENDFOR
		\ENDFOR
		\ENDFOR
	\end{algorithmic}
\end{algorithm}

\textbf{Actor:}
The actor learns the behavior policy and takes actions by employing a DNN with parameter $\theta_\mu^f$. 
The actor maximizes the following performance objective. 
\begin{equation} \begin{aligned} \label{actor performance f}
J^f(\theta_\mu^f)
&=\mathbb{E}_{s_t^f}[Q(s_t^f,a_t^f;\theta_Q^f)|_{a_t^f=\mu(s_t^f;\theta_\mu^f)}],
\end{aligned}\end{equation}
where $\mu(s_t;\theta_\mu)$ is a parameterized function that specifies the current policy by deterministically mapping the states to specific actions.
Following the analysis in \cite{DDPG}, the gradient of the current action policy can be written as 
\begin{equation} \label{actor gradient f}\begin{aligned}
\nabla_{\theta_\mu^f} J^f
=\mathbb{E}_{s_t^f\sim\mathcal{S}^f}&\Big[\nabla_{\theta_\mu^f}\mu(s_t^f;\theta_\mu^f) \nabla_{a^f}Q(s_t^f,a_t^f;\theta_Q^f)|_{a_t^f=\mu(s_t^f;\theta_\mu^f)}\Big].
\end{aligned}\end{equation}
Similar to the critic, the old parameters of the actor's target network are softly updated, and the evaluate network parameters $\theta_\mu^f$ are updated as  
\begin{equation}\label{actor update f}
\theta_\mu^f \leftarrow \theta_\mu^f - \eta_\mu^f \nabla_{\theta_\mu^f} J^f,
\end{equation}
where $\eta_\mu^f$ is the learning step size in the actor.

The pseudocode of the proposed multi-agent DDPG learning algorithm can be found in \textbf{Algorithm \ref{algo MADDPG}}, where the leader and the follower act sequentially and then train simultaneously.

\subsection{Upper Bound Analysis for MARL} \label{subsect upper bound}

The game-theoretic approach in Section \ref{subsect analysis cooperative} provides an upper-bound performance under the assumption of the ideal situation where the instantaneous CSI is available, and indicates the existence of a stable state for the individually rational agents after training. 
It helps confirm the stability of the MARL-based approach when the environment stops changing or changes very slowly. 
Specifically, MARL is a value-based learning process. 
Each agent learns its action policy through trial-and-error to find the best strategy $a_t=\mu(s_t)$ that can maximize the action-value function $Q(s_t, a_t)=\mathbb{E}[R_t|s_t,a_t]$.
When only the outdated CSI is available and the underlying transition probability of channel states is unknown, we have $s_{t+1}=H_t$, and 
\begin{equation} \begin{aligned} \label{action_value_comment13}
&{Q}(s_t, a_t) = \mathbb{E}\big[ R_t| s_t, a_t\big] \\
&=\int_{s_{t+1}} \!p(s_{t+1}|s_t,a_t) r(s_{t+1}, a_t) \mathrm{d} s_{t+1} \!+ \!\gamma \int_{s_{t+1}} \!p(s_{t+1}|s_t,a_t) \\
&
\times \int_{a_{t+1}} \pi(a_{t+1}|s_{t+1}) {Q}(s_{t+1}, a_{t+1}) \mathrm{d}a_{t+1} \mathrm{d}s_{t+1} \\
&= \mathbb{E}_{s_{t+1}}\big[r(s_{t+1}, a_t)\big] + \gamma\mathbb{E}_{s_{t+1}, a_{t+1}}\big[{Q}(s_{t+1}, a_{t+1})\big], 
\end{aligned}\end{equation}
where $r(s_{t+1}, a_t)$ represents the returned utility for the agent when the joint action $a_t=[a^f, a^l]$ is evaluated under $s_{t+1} = H_t$. 
Consider that the game is a one-step decision process, that is, ${Q}(s_t, a_t)$ is independent of ${Q}(s_{t+1}, a_{t+1})$.
Thus, the second term on the right-hand side (RHS) of (\ref{action_value_comment13}) is 0 and, subsequently, ${Q}(s_t, a_t) = \mathbb{E}_{s_{t+1}}\big[r(s_{t+1}, a_t)\big]$.

Let $a_t^*=\mu^*(s_t)$ be the optimal action policy of the individually rational RL-agent. We have
\begin{equation}\label{result_comment13_0}
    {Q}(s_t, a_t^*) = 
    \max_{a_t} {Q}(s_t, a_t)
    = \max_{a_t} \mathbb{E}_{s_{t+1}}\big[r(s_{t+1}, a_t)\big]. 
\end{equation}
The objective in (\ref{result_comment13_0}) shows that RL-agent's optimal action (derived from $s_t$) may not achieve the optimal utility with respect to each possible $s_{t+1}$, but can achieve the optimal average utility on any possible $s_{t+1}$.

On the other hand, when using game-theoretic approaches, we derive the optimal decisions $a^{f}=P_k^*$ and $a^{l}=C_k^*$ under the assumption of the instantaneous CSI, $H_t$. 
This optimal game-theoretic action can achieve the optimal utilities $U^r(a_t | H_t)$ and $U^s(a_t | H_t)$, under any possible $s_{t+1}$.

Therefore, we can establish the connection between the proposed game-theoretic and MARL-based approaches, i.e., 
\begin{align}
   \max_{a_t} U(a_t | H_t) = &\max_{a_t} r(s_{t+1}, a_t | a_t=\mu(s_{t+1})) \nonumber \\
   \triangleq & \tilde r(s_{t+1}), 
\end{align}
where $\tilde r(s_{t+1})$ represents the maximum utility under $s_{t+1}$. 
   
Note the optimal game-theoretic policy is derived based on $s_{t+1}$ (i.e., $H_t$), and the optimal RL policy is trained based on $s_t$ (i.e., $H_{t-1}$). 
When evaluating these two policies under any $s_{t+1}$ (i.e., $H_t$) and comparing their utilities, we have $\tilde r(s_{t+1}) \geq r(s_{t+1}, a_t^*|a_t^*=\mu^*(s_t))$. 
Moreover, we have 
\begin{align}\label{result_comment13}
     {Q}(s_t, a_t^*) = &\max_{a_t} \mathbb{E}_{s_{t+1}}\big[r(s_{t+1}, a_t | a_t=\mu(s_t))\big] \nonumber\\
     \leq & \mathbb{E}_{s_{t+1}}\big[ \tilde r(s_{t+1}) \big].
\end{align}
It can be concluded that the utilities achieved by the game-theoretic approach provide the upper bounds for those achieved by the MARL-based method.

\subsection{Complexity Analysis} \label{subsect complexity}
The time complexity and space complexity of the proposed MARL method can be estimated based on its neural network architecture. 
The time complexity is measured in the number of floating point operations per second (FLOPS). 
When calculating FLOPS, a FLOP accounts for a basic operation (e.g., addition, subtraction, multiplication, division, exponentiation, square root, etc.)~\cite{9580591, 8731635}.
On the other hand, space is needed to store network parameters and learning transitions.
The space complexity depends on the number of neurons in a network and the buffer size of the RL algorithm.
The complexity of the training and testing stages is discussed below. 

\textbf{Training: }
Recall that both agents have an actor and a critic part, and each part has two networks known as the evaluate and target networks, respectively. 
Different from image/video processing, the networks in both the actor and critic only have fully connected layers. 
Since all the networks used in the proposed method are isomorphic, we start by analyzing the evaluate network in the relay agent's actor part. 

For illustration convenience, suppose that the evaluate network has $M^{l,a}$ layers in the relay agent's actor part, and the $m^{l,a}$-th layer has $N_m^{l,a}$ neurons. 
Obviously, the first layer corresponds to the input layer of the network, and  $N_1^{l,a}=|\mathcal{S}_t^l|$. 
The last layer corresponds to the output layer of the network, and  $N_{M^{l,a}}^{l,a}=|\mathcal{A}_t^l|$. 
	Between the $m^{l,a}$-th and $(m+1)^{l,a}$-th layer, for each neuron in the $(m+1)^{l,a}$-th layer, there are three steps. 
	The first step is a dot product operation, which takes $N_m^{l,a}$ multiplications and $N_m^{l,a} -1$ additions. 
	The second step is a bias addition operation, which takes one addition.
	The third step is an activation operation, which generates the final output of the neuron.
 
	Since different types of activation functions have different amounts of basic operations, we use the variable $\kappa_{m+1}^{l,a}$ to denote the number of basic operations required by the activation function in the $(m+1)^{l,a}$-th layer.
	For each neuron in the $(m+1)^{l,a}$-th layer, there are a total of $2 N_m^{l,a} + \lambda_{m+1}^{l,a}$ basic operations. 
	Further, since there are $N_{m+1}^{l,a}$ neurons in the $(m+1)^{l,a}$-th layer, the time complexity between the two layers is $(2 N_m^{l,a} + \lambda_{m+1}^{l,a}) N_{m+1}^{l,a}$.
	For the actor network with $M^{l,a}$ layers, its training requires $\sum_{m=1}^{M^{l,a}-1}\big( (2 N_m^{l,a} + \lambda_{m+1}^{l,a}) N_{m+1}^{l,a} \big)$ FLOPs, and the time complexity is $\mathcal{O}(\sum_{m=1}^{M^{l,a}-1} N_m^{l,a} N_{m+1}^{l,a})$.

Likewise, suppose that the evaluate network has $M^{l,c}$ layers in the relay agent's critic part, and the $m^{l,c}$-th layer has $N_m^{l,c}$ neurons. 
The time complexity of training relay agent's critic network is $\mathcal{O}(\sum_{m=1}^{M^{l,c}-1} N_m^{l,c} N_{m+1}^{l,c})$.
As a result, the overall time complexity of training the relay agent is $\mathcal{O}(\sum_{m=1}^{M^{l,a}-1} N_m^{l,a} N_{m+1}^{l,a} + \sum_{m=1}^{M^{l,c}-1} N_m^{l,c} N_{m+1}^{l,c} )$

According to Section \ref{subsect DDPG}, the architecture of the source agent is the same as that of the relay agent, and the two agents are trained in a distributed manner. 
Suppose the total number of episodes $u_{max}^{epi}$ with $t_{max}$ epochs per episode, the overall time complexity of the training process in \textbf{Algorithm \ref{algo MADDPG}} is
\begin{equation} \label{time training}
\mathcal{O} \bigg( u_{max}^{epi} t_{max} (\sum_{m=1}^{M^{l,a}-1} N_m^{l,a} N_{m+1}^{l,a} + \sum_{m=1}^{M^{l,c}-1} N_m^{l,c} N_{m+1}^{l,c}) \bigg).
\end{equation}

The space that a fully connected layer needs is to store a weight matrix and a bias vector. 
In addition, the experience relay buffer needs space to store experiences for training. 
Because each agent consists of four neural networks and one experience relay buffer, the space complexity of the training process in \textbf{Algorithm \ref{algo MADDPG}} is given by
\begin{equation} \begin{aligned} \label{space training}
\mathcal{O} &\left(2\bigg[ 2 \sum_{m=1}^{M^{l,a}-1}\big( N_m^{l,a} N_{m+1}^{l,a} + N_{m+1}^{l,a} \big) + \right.\\
&\left.2 \sum_{m=1}^{M^{l,c}-1}\big( N_m^{l,c} N_{m+1}^{l,a} + N_{m+1}^{l,c} \big) + 
|\mathcal{B}^l| \bigg]\right) \\
=& \mathcal{O} \bigg( \sum_{m=1}^{M^{l,a}-1} N_m^{l,a} N_{m+1}^{l,a} + \sum_{m=1}^{M^{l,c}-1} N_m^{l,c} N_{m+1}^{l,a} + N_{m+1}^{l,c} + |\mathcal{B}^l| \bigg).
\end{aligned}\end{equation}

\textbf{Testing: }
In the testing stage, the time complexity depends primarily on the policy selection process.
Since the critic is designed to help the actor learn its action policy during training, it does not operate during testing. 
Therefore, we only need to focus on the architecture of the actor.
Based on the analysis above, we can directly obtain the time complexity of the distributed execution in the testing stage as
\begin{equation} \begin{aligned} \label{time test}
\mathcal{O} \bigg(\sum\nolimits_{m=1}^{M^{l,a}-1} N_m^{l,a} N_{m+1}^{l,a} \bigg).
\end{aligned} \end{equation}

The buffers do not function in the testing stage. 
Therefore, the space complexity of the distributed execution is given by
\begin{equation} \label{space test}
\mathcal{O} \bigg(\sum\nolimits_{m=1}^{M^{l,a}-1} N_m^{l,a} N_{m+1}^{l,a} \bigg).
\end{equation}

\section{Numerical Evaluation}\label{sect Result}
In this section, we introduce the setup of experiments and then evaluate the performance of our proposed methods.

\subsection{Experiment Setup}
We set the channel state to change according to  (\ref{channel state}) with parameter $\rho=0.8$. 
The transmit power of the source is $P_s=1.0$ W. The maximum available power of each relay is $P_{max}=2.0$ W. 
The maximum price for per unit of power is $C_{max}=10.0 /$ W. 
We set $P_{min}=0$ and $ C_{min}=0$. 
The learning rates for each actor and critic are $\eta_\mu^f=\eta_\mu^l=0.001$ and $\eta_Q^f=\eta_Q^l=0.005$, respectively. 
The soft update parameter is $\tau=0.001$. 
Similar to the settings in \cite{9453731}, we set the buffer size to $|B^f|=|B^l|=10000$, and the mini-batch size for learning to 128. 
In addition, the parameters in reward functions are set to $\alpha=\beta=0.1$.
The source code is available at: https://github.com/gyz1997/MARL-for-Stackelberg-game.git.

For comparison, the following benchmarks are considered.
\begin{itemize}
\item 
\textbf{Game-Based Solution (GBS): }
This is a typical algorithm for solving games. Similar to \cite{9444840} and \cite{9685510}, we adapt the game-based algorithm to the problem studied. Assuming that agents' utility functions are known, and the instantaneous CSI is available, following the game theory analysis in Section-\ref{game analysis}, we can obtain the result that each agent is considered rational and intelligent enough. Therefore, such results can, to some extent, be regarded as the optimal solution.
	
\item
\textbf{DQN-MARL: } 
This algorithm is extended from the multi-agent DQN method originally designed in \cite{10056414, 10285368} for cooperative learning with partial observation. 
Multi-agent DQN is widely adopted to solve game problems.
Since DQN requires a discrete action space, we discretize the power price and power in our experiments.

\item
\textbf{Learning and Gaming Mixed Solution (LGMS): } 
This algorithm is extended from the DDPG-based methods originally designed in \cite{9453731, 10101816} for communication network optimization in a centralized control manner. 	
Based on the CSI of the previous time slot, the relay agent performs actions according to the policy trained by the DDPG method. 
Then, based on outdated CSI and the relay agent's policy, the source agent directly derives its solution using the game-based method.

\item
\textbf{MARL for Distributed Cooperative Relays (RLDCR): }
This algorithm is extended from the DDPG-based methods originally designed in \cite{9453731, 10101816} for communication network optimization in a centralized manner. 
Different from the proposed method, each relay is an independent agent in RLDCR, and performs actions based on local observations.
The agents vote for one relay to forward packets for the source, and also provide reference prices.
The final decision is made through the voting.

\item
\textbf{Random: }
Both the relay and source agents perform actions randomly; that is, the relay alliance randomly chooses a relay and sets a random price for power, while the source determines the amount of power randomly for forwarding its signals.

\end{itemize}

\subsection{Numerical Results}
\begin{figure}
    \centering
    \vspace{-0.35cm}
    \subfigtopskip=1pt 
    \subfigbottomskip=1pt
    \subfigcapskip=-4pt 
    
    \subfigure[ ]{
		\label{different scenario utility}
		\includegraphics[scale=0.55]{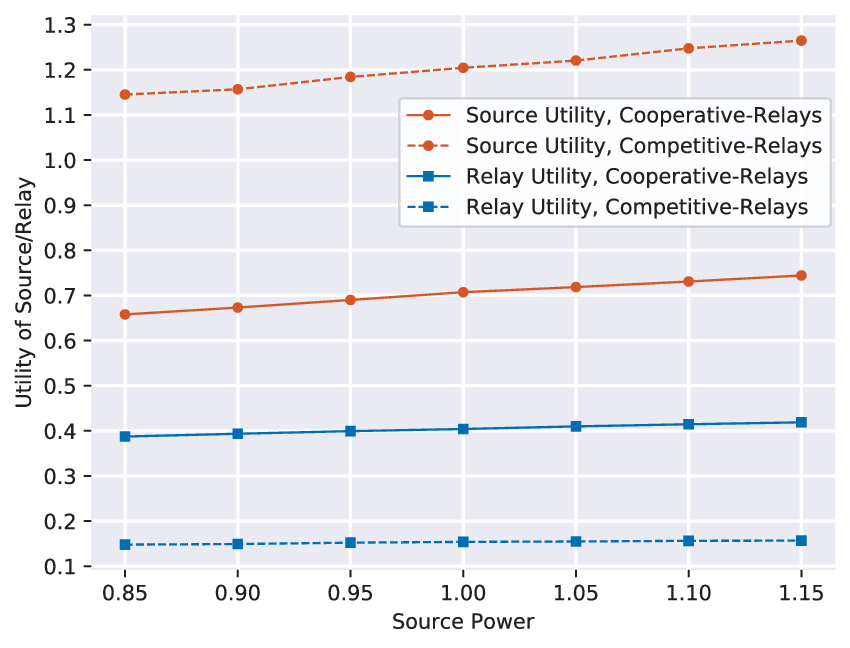}
	}
    \subfigure[ ]{
		\label{different scenario capacity}
		\includegraphics[scale=0.55]{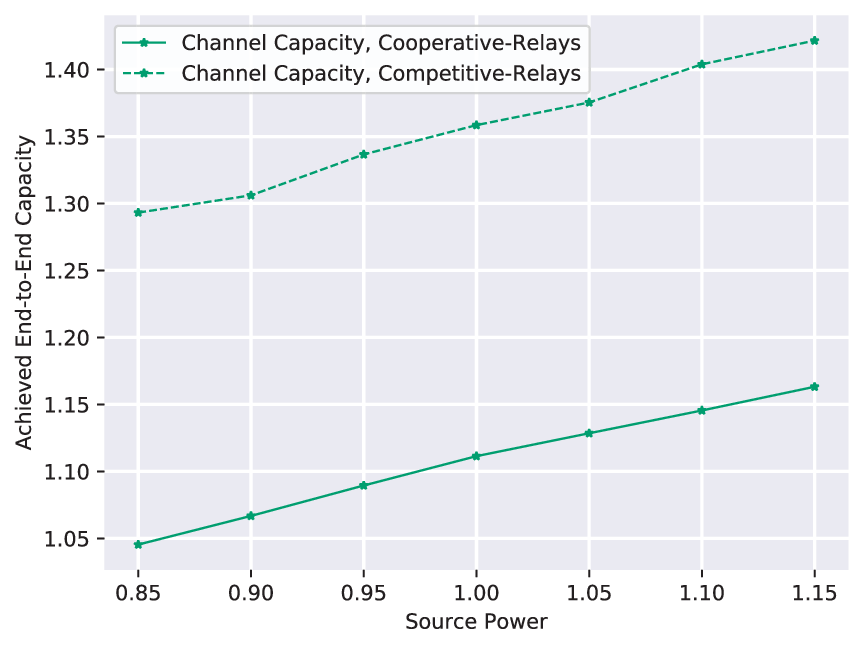}
	}

    \caption{\small Performance comparison between the different relay settings under different levels of the transmit power at the source.}
    \label{different scenario}
\end{figure}

To validate \textbf{Proposition 4}, we use the game-theoretic approach to evaluate the performance of the source and the relays under the setting of competitive relays, and compare it with the proposed cooperative setting of a relay alliance. 
The results are depicted in Fig. \ref{different scenario}. 
The solid lines indicate the utilities and the achieved channel capacity under the setting of a relay alliance. The dashed lines indicate those under the setting of competitive relays.  

Fig. \ref{different scenario utility} plots the utilities of the source and relay, as the transmit power of the source, $P_s$, increases from 0.85 W to 1.15 W. 
The utilities of all players increase with the source's transmit power. 
The source can achieve higher utilities under the competitive setting than under the relay alliance. 
However, the utilities of the relay alliance under the cooperative setting are higher than those under the competitive setting. 
Take $P_s=1.0$ as an example.
The utilities of the source and relay under the cooperative setting are about 0.707 and 0.404, respectively. 
By contrast, under the competitive setting, the utilities of the source and relay are about 1.204 and 0.154, respectively.
The simulation results are consistent with the analysis in \textbf{Proposition 4}.

Fig. \ref{different scenario capacity} depicts the achieved end-to-end channel capacity, as $P_s$ increases. 
Take $P_s=1.0$ as the example.
The achieved end-to-end channel capacity is 1.11 bits/s/Hz under the cooperative setting and 1.34 bits/s/Hz under the competitive setting. 
The channel capacity is much higher in the competitive setting than in the cooperative setting.
This is consistent with the utility of the source shown in Fig. \ref{different scenario utility}, because the relays have to compete even by sacrificing their own interest (i.e., utility $U^r$). 
On the contrary, the source can take advantage of the competition between the relays, achieving higher utilities and end-to-end channel capacity. 
Such ill competition is prevented under the cooperative setting by creating the relay alliance. 
The relays can therefore obtain higher revenue. 

\begin{figure}[htb]
	\centering
	
	\vspace{-0.35cm} 
	\subfigtopskip=1pt 
	\subfigbottomskip=1pt 
	\subfigcapskip=-4pt 
	
	\subfigure[ ]{
		\label{train_utility_source}
		\includegraphics[scale=0.55]{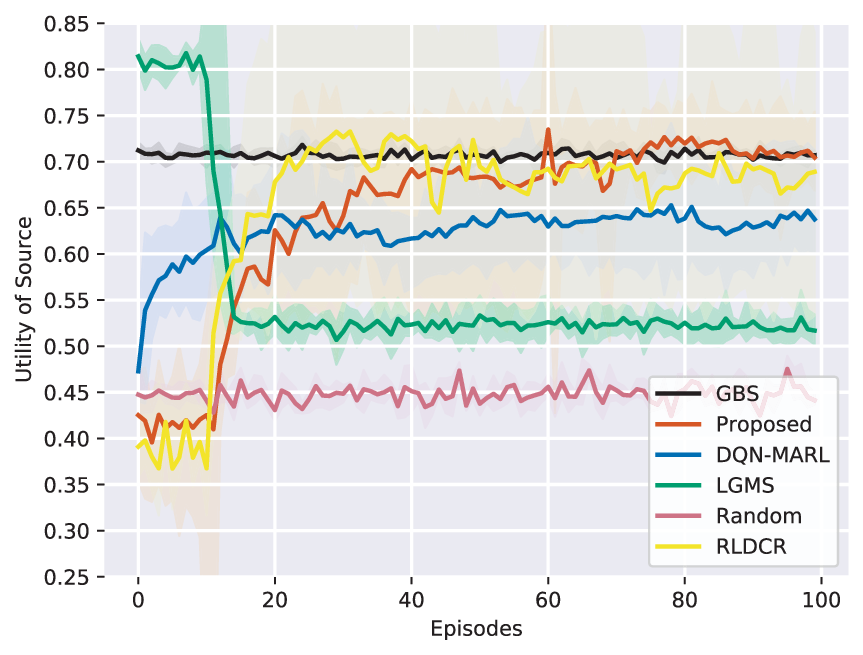}
	}
	\\ 
	\subfigure[ ]{
		\label{train_utility_relay}
		\includegraphics[scale=0.55]{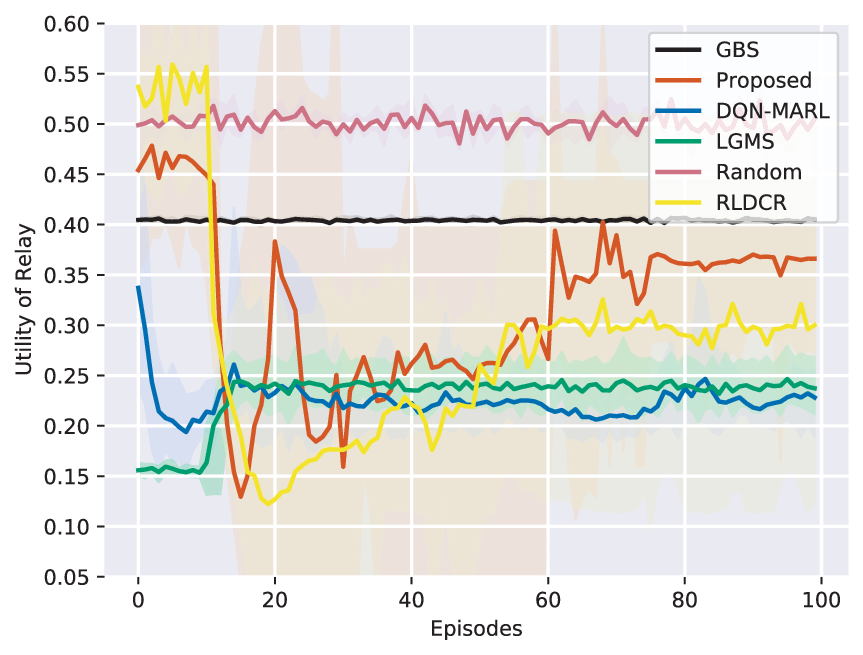}
	}
	\\

	\subfigure[ ]{
		\label{train_channel_capacity}
		\includegraphics[scale=0.55]{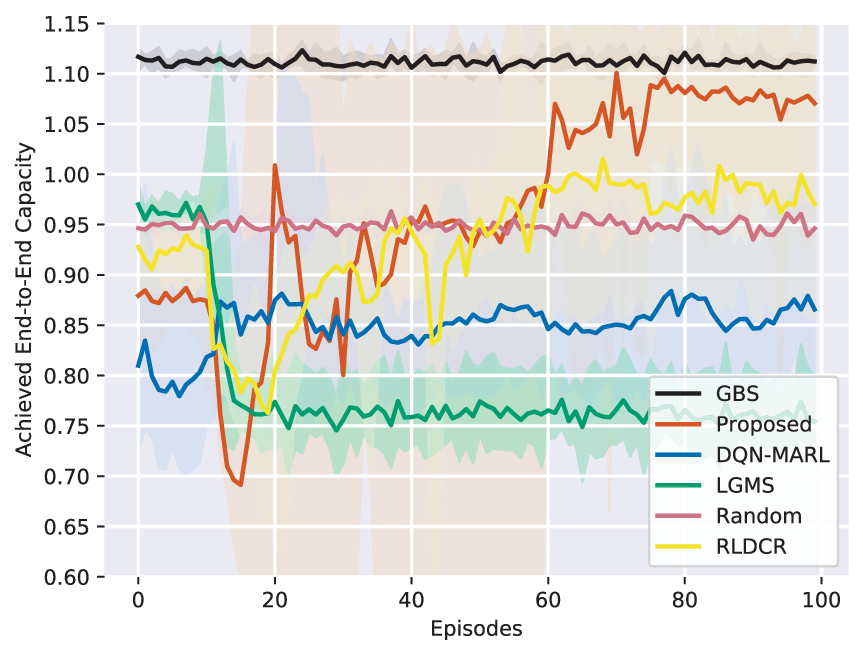}
	}
	
	\caption{Performance comparison between the different methods in the training stage.}
	\label{train}
\end{figure}

As discussed, the relays are more likely to choose to cooperate and form an alliance to compete with the source to increase their revenue. 
In the rest of this section, we concentrate on the cooperative setting of a relay alliance.
We proceed to evaluate the different methods in the training stage with the total number of episodes $t^{epi}_{max}=100$ and the maximum number of time slots per episode $t_{max}=1000$.
The training results are depicted in Fig. \ref{train}, where the dark line represents the median of 10 successful trails, and the shadow region represents their range.

Figs. \ref{train_utility_source} and \ref{train_utility_relay} plot the utilities of the source and relay agent with the increase of training episodes, respectively. 
Firstly, using the GBS method, we can obtain the upper bound performance of the source's and relay's utilities, which are about 0.707 and 0.404, respectively. 
The achieved performance is better than that of the other learning methods because it is assumed that the game players' optimization goals are known and the instantaneous CSI is available. 
It is also observed that during the process of gaming and learning, 
the proposed method and the benchmarks, including LGMS and DQN-MARL, can eventually converge to a stable value.
The agents in multi-agent learning must undergo policy testing and adjust before achieving a policy close to the equilibrium.
In the last few episodes, the agents successfully obtain policies that constitute the equilibrium.

When using the proposed method and LGMS, the first ten episodes collect experience units, and thus, random selection is made until the buffer is filled. 
It is found that, after performing policy training, the LGMS can converge faster and achieve better training stability than our proposed method. 
Specifically, the LGMS converges after about the 15th episode, while the proposed method converges after about the 75th episode. 
Moreover, the proposed algorithm undergoes stronger fluctuation during training.
The reason is that the LGMS is a single-agent method, which does not suffer from the non-stationary problem of multi-agent learning and can quickly converge to a stable policy.

Nevertheless, the proposed method can substantially outperform the LGMS method.
The utilities of the source and relay using the proposed method are about 0.706 and 0.373, respectively, while those achieved by the LGMS are about 0.524 and 0.247. 
The reason is that the agents' decisions are based on outdated CSI, which sometimes can mislead the agents' actions. 
When using the LGMS method, the source's solution is directly derived from game theory using outdated CSI, which can hardly be optimal. 
On the contrary, our proposed method equips learning algorithms for both agents, improving the rationality of both agents' actions.

We can find that the DQN-MARL method also converges faster but performs worse than our method. 
The DQN-MARL converges after about the 50th episode with slight fluctuation, and the achieved utilities of source and relay are about 0.646 and 0.231, respectively. 
This is because the action space for the DQN agent is discretized. Thus, the policy space for the game is finite, making it easier for the agents to find a stable solution. 
On the other hand, the discretized action space can introduce errors. 
A non-optimal policy of an agent may cause the action policy of the other agent to deviate further from the optimal solution. 
By contrast, our proposed method searches for the equilibrium solution in a continuous policy space, which converges slower but has a better chance of finding an optimal solution. 
As a result, the utilities achieved by the proposed method are closer to those of the GBS compared to the three methods that use the learning algorithm.
Given the difficulty in obtaining instantaneous CSI and the agents' utility goals can hardly be known to their opponents, a small performance gap between the optimal solution and ours is acceptable to a certain extent.

    The RLDCR method requires each agent to learn in parallel, instead of forming an alliance to learn on behalf of the relays.  
    The first ten training episodes collect experience units, and random selection is performed. 
    It is observed that the proposed method outperforms RLDCR. 
    The utilities of the source and relays are about 0.681 and 0.302, respectively, under the RLDCR method. 
    The reason is that RLDCR makes each relay an agent learn based on its local observation, increasing the number of agents in the environment and, accordingly, the non-stationarity of the MARL system.
    In addition, the architecture of distributed learning and action among the cooperative relays may suffer from the lazy agent problem \cite{sunehag2017value, foerster2018counterfactual, rashid2020monotonic}.
    When some of the cooperative agents in the team have learned good strategies, the other agents can receive good team rewards without doing anything. 
    By contrast, our method considers a relay alliance that operates centrally and avoids the lazy agent problem to the greatest extent.

\begin{figure} 
	\centering
	
	\vspace{-0.35cm} 
	\subfigtopskip=1pt 
	\subfigbottomskip=1pt 
	\subfigcapskip=-4pt 
	
	\subfigure[ ]{
		\label{test1_utility_source}
		\includegraphics[scale=0.55]{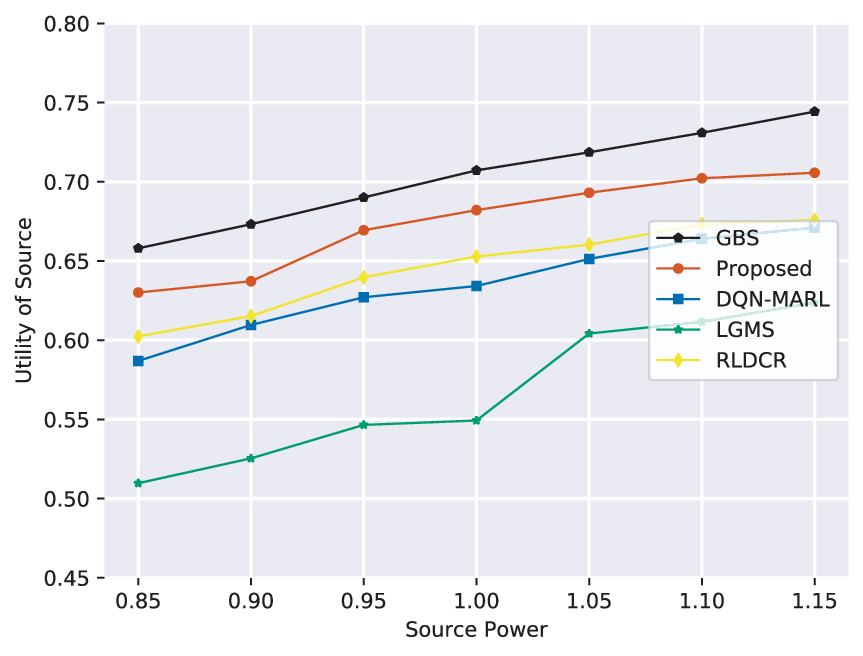}
	}
	\\
	\subfigure[ ]{
		\label{test1_utility_relay}
		\includegraphics[scale=0.55]{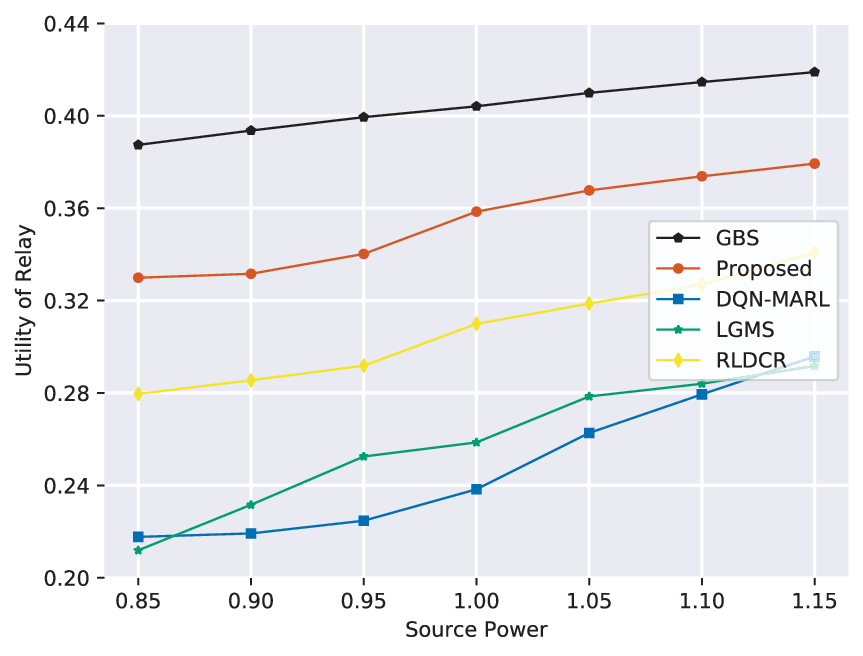}
	}
	\\
	\subfigure[ ]{
		\label{test1_channel_capacity}
		\includegraphics[scale=0.55]{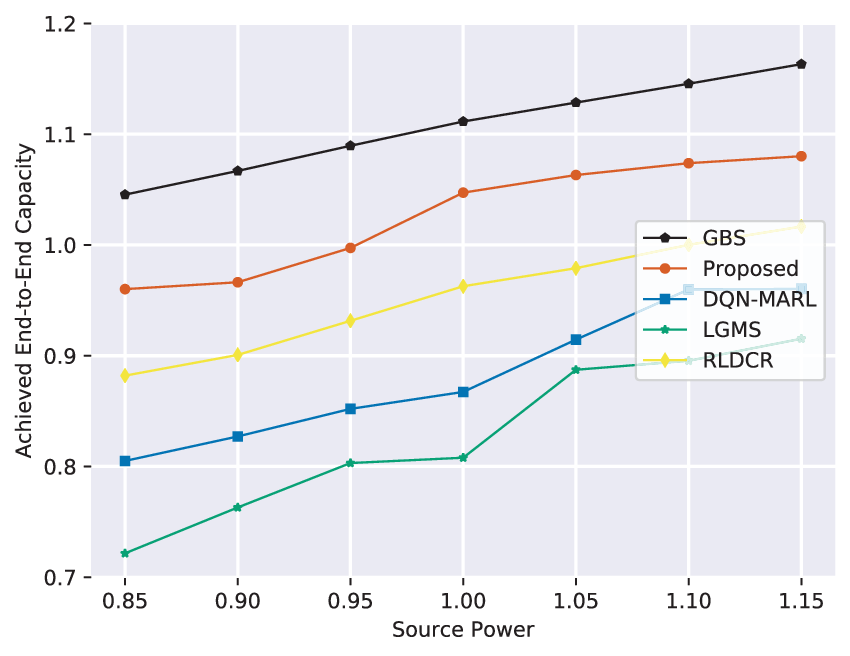}
	}
	
	\caption{Performance of different methods under different levels of the transmit power at the source.}
	\label{test_source_power}
\end{figure}
It is obvious that the Random method benefits some players at the expense of others.
In situations where all players are rational, such an approach is ineffective.
Fig. \ref{train_channel_capacity} plots the achieved end-to-end channel capacity between source and destination with increased training episodes. 
Similarly, the GBS method can achieve the upper bound performance of about 1.111 bits/s/Hz.
When using our proposed method, we can obtain the best performance among the other methods. 
The achieved channel capacity is about 1.079 bits/s/Hz, only about 2.9\% away from the optimum.
The performances of DQN-MARL, LGMS, RLDCR, and Random method are not good enough, which are only about 0.877, 0.771, 0.980, and 0.951 bit/s/Hz, respectively. 

Next, we test the above methods by changing the hyper-parameters in the game problem. 
In Fig.~\ref{test_source_power}, we adjust the source power $P_s$ ranging from 0.85 to 1.15, and compare our proposed method with the other four game policies, which are the GBS, DQN-MARL, LGMS, and RLDCR methods. 
Figs. \ref{test1_utility_source} and \ref{test1_utility_relay} plot the utilities of the source and relay with the increase of source power, respectively. 
It can be observed that the utility curves of all the methods increase with the increase of source power.
Among all the methods, the GBS method achieves the best result, which also represents the upper-bound performance of the investigated game. 
Take $P_s=1.0$ W as an example. Since the GBS method does not involve the training of DNNs, its performance in the test stage is similar to that in the training stage. 
The utility of the source and relay achieved by our method are about 0.682 and 0.359, respectively. 
It is a little worse than those obtained in the training stage because the networks are fixed and do not have further training in the testing stage. 
However, the achieved performance of our method is still better than those using the DQN-MARL, LGMS, and RLDCR methods. 
The source utility of the DQN-MARL, LGMS, and RLDCR methods are about 0.634, 0.549, and 0.653, respectively.
The relay utility of the DQN-MARL, LGMS, and RLDCR methods are about 0.238, 0.259, and 0.310, respectively. 

Fig. \ref{test1_channel_capacity} depicts the achieved end-to-end channel capacity with increased source power. 
Take $P_s=1.0$ as the example, again. 
The results of the GBS, Proposed, DQN-MARL, LGMS and RLDCR methods are about 1.11, 1.05, 0.86, 0.81 and 0.96 bit/s/Hz, respectively. 
The performance of our proposed method is about 5.4\% worse than that of the optimum but has an improvement of about 18\%, 22\% and 9\%, compared to the DQN-MARL, LGMS and RLDCR, respectively. 
As mentioned earlier, the performance gap between ours and the optimum is acceptable because part of the information in the game is difficult to obtain. 
Our approach can improve the utility of the whole system, compared to the other learning methods.

\begin{figure} 
	\centering
	
	\vspace{-0.35cm} 
	\subfigtopskip=1pt 
	\subfigbottomskip=1pt 
	\subfigcapskip=-4pt 
	
	\subfigure[ ]{
		\label{test2_utility_source}
		\includegraphics[scale=0.55]{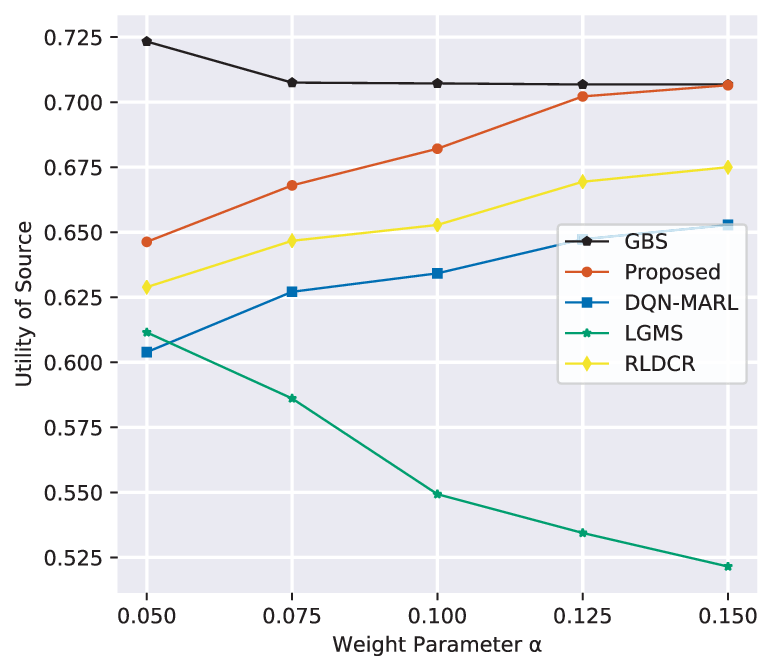}
	}
\\
	\subfigure[ ]{
		\label{test2_utility_relay}
		\includegraphics[scale=0.55]{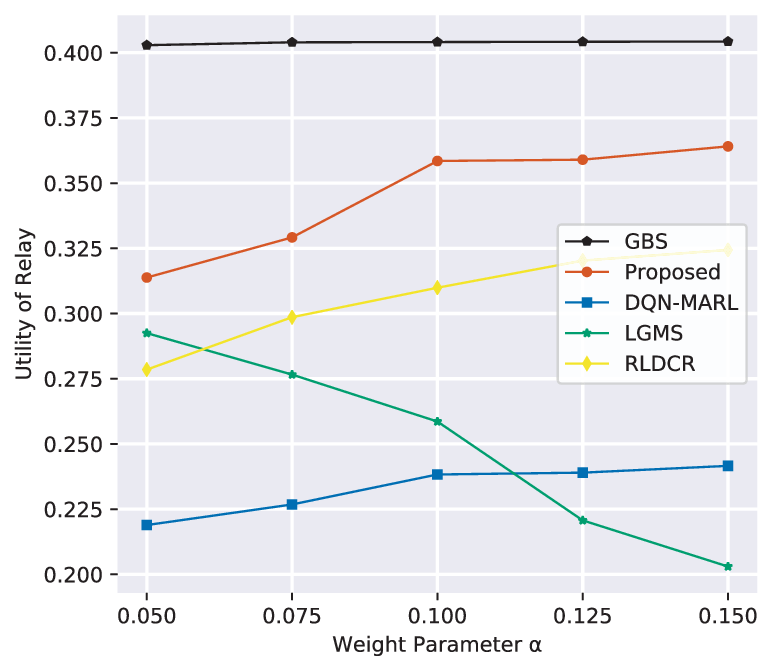}
	}
\\
	\subfigure[ ]{
		\label{test2_channel_capacity}
		\includegraphics[scale=0.55]{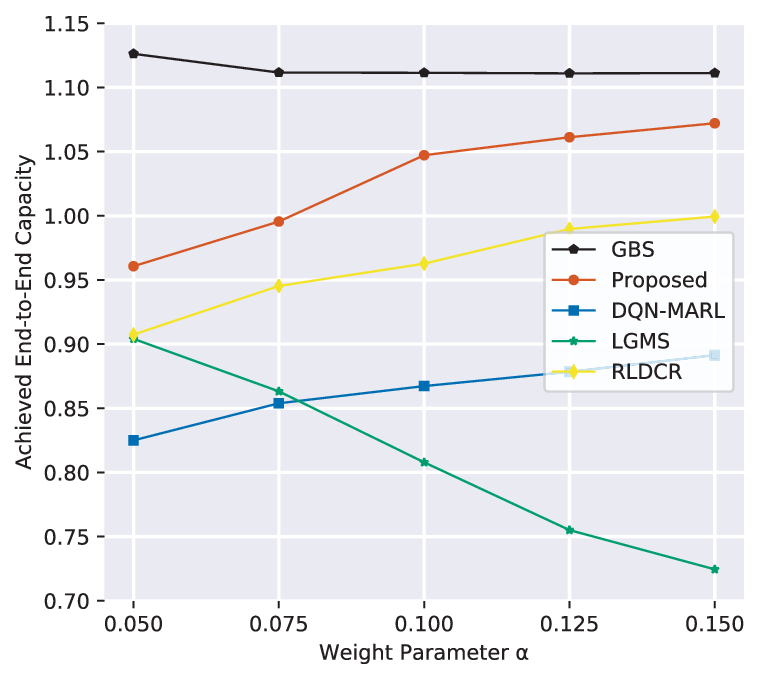}
	}
	
	\caption{Performance of different methods under different utility weighting parameters.}
	\label{test_weight_parameter}
\end{figure}

In addition, we evaluate the performance of the methods by adjusting the weight parameter $\alpha$ in utility functions ranging from 0.05 to 0.15. $\beta=\alpha$. 
The results can be found in Fig. \ref{test_weight_parameter}.
The increase of the weight parameter in utility functions will make the source pay more attention to its power cost. 
By jointly inspecting all subfigures in Fig. \ref{test_weight_parameter}, little difference is noticed between the results of the GBS method under different utility weighting parameters. 
When using learning-based methods, the curves of the Proposed, DQN-MARL and RLDCR methods exhibit the same trend.
Their obtained source utility, relay utility, and achieved channel capacity all increase as the utility weighting parameter increases. 
Take $\alpha=0.15$ as an example. The optimal equilibrium utilities for the source and relay are 0.707 and 0.404, respectively. 
The source utility of the Proposed, DQN-MARL, and RLDCR methods are about 0.707, 0.653, and 0.675, respectively. 
The relay utility of the Proposed, DQN-MARL, and RLDCR methods are about 0.364, 0.242, and 0.324, respectively.
The performance of our method is a little worse than the optimum obtained by the GBS, but obviously outperforms the DQN-MARL and RLDCR methods.
When it comes to the achieved end-to-end channel capacity, as shown in Fig. \ref{test2_channel_capacity}, the performance of the GBS, Proposed, DQN-MARL and RLDCR methods are about 1.11, 1.07, 0.89 and 1.00 bit/s/Hz, respectively.
Our method is about 3.7\% worse than the optimum, but 16\% and 7\% better than the DQN-MARL and RLDCR, respectively. 

Another interesting observation in Fig. \ref{test_weight_parameter} is that the variation trend of the LGBS method is completely opposite to the other learning methods. 
The larger the utility weight parameter, the worse the performance of each agent, primarily caused by outdated CSI.  
Since the source uses solutions derived from game theory under outdated CSI, its policy will have certain deviations compared to the optimal equilibrium solution, which also affects the policy learning of the relay agent.
When the utility weight parameter becomes larger, the impact of each agent's decision on the utility of each other also increases. 
These experimental results reflect the shortcomings of the LGBS method when applied to multi-agent tasks, and an in-depth cause analysis requires further research.

To sum up, the proposed method does not require instantaneous CSI, can achieve an acceptable performance that is close to the equilibrium, and outperforms single-agent LGMS, multi-agent DQN-MARL, and RLDCR.

\section{Conclusion}\label{sect Conclusion}
In this paper, we integrated MARL into a two-hop cooperative network to balance system performance and cost. 
Unlike traditional works where central control is typically implicitly considered, we established a non-cooperative Stackelberg game in which the source and relays have conflicting goals. 
We proved the existence of the equilibrium in the ideal scenario where the agents know each other's goal and instantaneous CSI. 
Then, we designed a DDPG-based MARL framework, which takes the source and the relay alliance as two agents. The agents do not have instantaneous CSI or each other's goal. 
Simulations revealed that our approach substantially outperforms the alternative learning methods. 
It can obtain solutions near the game-theoretic equilibrium for all players in the game, which is only 2.9\% away from the optimal solution available only in the ideal situation.

\section*{Acknowledgement}
The authors thank Lingzhi Xia, Shaoyi Han and Pengcheng Sun, from the College of Electronics and Information Engineering, Tongji University, for the insightful discussions and feedback on this paper. 
The authors would also like to thank their colleagues at the Sino-German Center of Intelligent Systems, Tongji University.

\bibliographystyle{IEEEtran}
\bibliography{paper_ref}

\end{document}